\documentclass[twocolumn,pre,superscriptaddress,showpacs]{revtex4}
\usepackage{graphicx}
\usepackage{amsmath}
\usepackage{amssymb}

\newcommand{\species}[1]{{\it #1}}
\newcommand{\Ecoli}{\species{E. coli}}

\newcommand{\TF}[1]{{\mathrm{#1}}}
\newcommand{\TFA}{\TF{A}}
\newcommand{\TFB}{\TF{B}}

\newcommand{\messenger}[1]{\mathrm{M}_{#1}}
\newcommand{\MA}{\messenger{\TFA}}
\newcommand{\MB}{\messenger{\TFB}}

\newcommand{\NA}{N_{\TFA}}
\newcommand{\NB}{N_{\TFB}}

\newcommand{\OO}{{\mathrm{O}}}
\newcommand{\OA}{{\mathrm{O}}_{\TFA}}
\newcommand{\OB}{{\mathrm{O}}_{\TFB}}

\newcommand{\nothing}{\emptyset}
\newcommand{\Nmean}{\overline N}

\newcommand{\kf}{k_{\mathrm{f}}}
\newcommand{\kb}{k_{\mathrm{b}}}
\newcommand{\kon}{k_{\mathrm{on}}}
\newcommand{\koff}{k_{\mathrm{off}}}
\newcommand{\kA}{k_{\TFA}}
\newcommand{\kB}{k_{\TFB}}
\newcommand{\rA}{r_{\TFA}}
\newcommand{\rB}{r_{\TFB}}
\newcommand{\ktA}{k''_\TFA}
\newcommand{\ktB}{k''_\TFB}
\newcommand{\kdA}{k'_\TFA}
\newcommand{\kdB}{k'_\TFB}

\newcommand{\tauescape}{\tau_{\leftrightarrow}}

\newcommand{\mudA}{\mu'_\TFA}
\newcommand{\mudB}{\mu'_\TFB}
\newcommand{\muA}{\mu_{\TFA}}
\newcommand{\muB}{\mu_{\TFB}}
\newcommand{\tmu}{\tilde\mu}
\newcommand{\tmuA}{\tmu_{\TFA}}
\newcommand{\tmuB}{\tmu_{\TFB}}
\newcommand{\Kd}{K_{\mathrm{d}}}
\newcommand{\Kb}{K_{\mathrm{b}}}
\newcommand{\Vcell}{V_{\mathrm{c}}}
\newcommand{\kVcell}{k\Vcell}
\newcommand{\lphage}{$\lambda$-phage}
\newcommand{\Prob}{\mathrm{Prob}}
\newcommand{\tick}{1}
\newcommand{\cross}{0}
\newcommand{\latin}[1]{\textit{#1}}
\newcommand{\PS}{P_{\mathrm{S}}}
\newcommand{\num}[1]{n_{#1}}

\begin{document}

\title{Chemical models of genetic toggle switches}

\author{Patrick B. Warren}
\affiliation{Unilever R\&D Port Sunlight,
Bebington, Wirral, CH63 3JW, United Kingdom}
\affiliation{FOM Institute for Atomic and Molecular Physics,
Kruislaan 407, 1098 SJ Amsterdam, The Netherlands.}

\author{Pieter Rein ten Wolde}
\affiliation{FOM Institute for Atomic and Molecular Physics,
Kruislaan 407, 1098 SJ Amsterdam, The Netherlands.}
\affiliation{Division of Physics and Astronomy, Vrije Universiteit, De
Boelelaan 1081, 1081 HV Amsterdam, The Netherlands}

\date{September 30, 2004}

\begin{abstract}
We study by mean-field analysis and stochastic simulations chemical
models for genetic toggle switches formed from pairs of genes that
mutually repress each other.  In order to determine the stability of
the genetic switches, we make a connection with reactive flux theory
and transition state theory. The switch stability is characterised by
a well defined lifetime $\tau$.  We find that $\tau$ grows
exponentially with the mean number $\Nmean$ of transcription factor
molecules involved in the switching.  In the regime accessible to
direct numerical simulations, the growth law is well characterised by
$\tau\sim\Nmean{}^{\alpha}\!\exp(b\Nmean)$, where $\alpha$ and $b$ are
parameters.  The switch stability is decreased by phenomena that
increase the noise in gene expression, such as the production of
multiple copies of a protein from a single mRNA transcript (shot
noise), and fluctuations in the number of proteins produced per
transcript.  However, robustness against biochemical noise can be
drastically enhanced by arranging the transcription factor binding
domains on the DNA such that competing transcription factors mutually
exclude each other on the DNA. We also elucidate the origin of the
enhanced stability of the exclusive switch with respect to that of the
general switch: while the kinetic prefactor is roughly the same for
both switches, the `barrier' for flipping the switch is significantly
higher for the exclusive switch than for the general switch.
\end{abstract}



\maketitle

\section{Introduction}
In an organism, genes can be turned on or off by the binding of
proteins to regulatory sites on the DNA in the vicinity of the
starting point for transcription \cite{gensigbook}.  The proteins are
known as transcription factors and the DNA binding sites are known as
operators.  The process is an example of gene regulation.
Transcription factors can turn genes off by stereochemical blockage of
the binding of RNA polymerase (RNAp), or they can turn genes on by
co-operative binding (recruitment) of RNAp \cite{gensigbook}.

Since transcription factors are proteins, they are coded for elsewhere
on the genome.  This means that transcription factors can regulate the
production of other transcription factors, or indeed can regulate
their own production.  A highly complex network of biochemical
reactions can be built up, capable, in principle, of solving
arbitrarily complex computational problems \cite{MOM,EPDM}.  The
network is interfaced to the outside world by, e.g., signalling
cascades from receptor proteins or by specific interactions between
transcription factors and small molecules such as metabolites.  In
this way, even relatively simple organisms such as \Ecoli\ can perform
fairly complex computations such as integrating different environmental
signals. In higher organisms, gene regulation networks lie at the
heart of cell differentiation and developmental pathways.

\begin{figure}
\begin{center}
\includegraphics{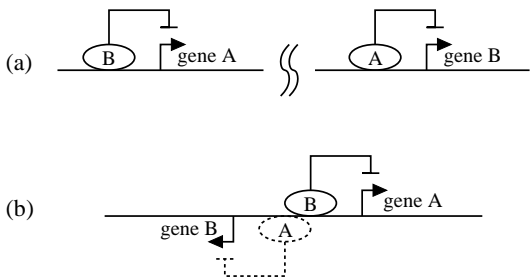}
\end{center}
\caption{(a) A genetic toggle switch can be formed from a pair of
genes that mutually repress each other's expression.  (b) If the genes are
transcribed on opposite strands of the DNA, the upstream regulatory
domains can overlap and interfere with one another. As a result,
competing regulatory molecules mutually exclude each other.\label{fig:switch}}

\end{figure}

Genetic toggle switches are an informative paradigm in this context
\cite{lphagebook,Ferrell}.  They are regulatory constructs which
select between two possible stable states, representing for example
differentiation between two developmental pathways.  Perhaps the
simplest kind of switch is one that is constructed from a pair of
genes that mutually repress each other, as indicated in
Fig.~\ref{fig:switch}.  The switch of \lphage\ in \Ecoli\ is a
naturally occuring example, which has been studied in much detail
\cite{lphagebook, ARMcA, AS, ABJS}. Another well-known example is that
of the human herpesvirus 3 (chickenpox or varicella-zoster virus),
which has a pathogenesis that is remarkably similar to \lphage;  the
virus lies dormant after the initial infection, but can be triggered
to re-emerge much later as shingles (herpes zoster).  Synthetic toggle
switches have also been constructed {\em in vivo} \cite{GCC}.

Genetic switches are usually flipped by external signals. In the
\lphage\ , for example, the switch is initially in the dormant
(lysogenic) state but can be flipped into the active (lytic) state by
the presence of the bacterial protein RecA.  Such an induction event
occurs when the cell starts to produce RecA to repair DNA damage as a
result of, e.g., a burst of UV light.

Importantly, genetic switches are often so stable that they remain in
one state until the external trigger flips the switch.  In wild-type
\lphage\ spontaneous flips are extremely rare, occurring at a rate of
perhaps as low as one spontaneous flip in $10^{12}\,\mathrm{s}$
\cite{ABJS}. An important question, therefore, is: what are the design
principles that allow the switch to attain such extreme stability in the
presence of fluctuations and biochemical noise?  This question is
particularly relevant, because detailed modelling has suggested that
the stability of \lphage\ cannot be explained on the basis of
our current understanding \cite{ABJS}.

We have recently shown that mutual exclusion of competing
transcription factors can drastically enhance the stability of genetic
switches \cite{WtW04_1}. Transcription factors can mutually exclude each
other, if their operator sites (partially) overlap (see
Fig.~\ref{fig:switch}b). It appears that nature has exploited the
functional benefit of this spatial arrangement of operators, since a
recent statistical analysis has revealed that this network motif of
`overlapping operons' is over-represented in the bacterium \Ecoli\
\cite{WtW04_2}.

In this paper, we extend the analysis of Ref. \onlinecite{WtW04_1} in
several ways \cite{note4}. In the next section, we first describe the
set of chemical reactions by which we model the genetic toggle
switches (see Fig.~\ref{fig:switch}).  We then turn to a mean-field
(chemical rate equation) analysis of the appearance of switching
behaviour. This mean-field analysis shows that the region of
bistability is significantly larger for the exclusive switch than for
the general switch.  In order to determine the life times of the
stable steady states of the switches, we have performed direct
stochastic simulations; these are described in
section~\ref{sec:simulations}. We then consider a number of
refinements of the basic switch model.  In particular, we explicitly
take into account transcription and translation. We show that both
`shot noise' and fluctuations in the number of proteins produced per
mRNA transcript markedly decrease the switch stability. However, we
again find that the exclusive switch is still orders of magnitude more
stable than the general switch. In this paper, we also elucidate the
origin of the enhanced stability of the exclusive switch with respect
to that of the general switch: while the kinetic prefactor is roughly
the same for both switches, the `free-energy' barrier for flipping is
significantly higher for the exclusive switch  than for
the general switch \cite{WtW04_1}. This underscores our earlier observation that
mutual exclusion can drastically enhance the robustness of genetic
switches to biochemical noise \cite{WtW04_1}. It also strongly
supports the conjecture that regulatory control is one of the main
evolutionary driving forces that shape the organisation of genes and
operons on the genome \cite{WtW04_2}.

\section{Model specification}
Our starting point is a set of chemical reactions which model the
processes involved in the toggle switches shown in
Fig.~\ref{fig:switch}.  As chemical species, we introduce a pair of
transcription factors (TFs) which can exist as monomers, $\TFA$ and
$\TFB$, or multimers, $\TFA_n$ and $\TFB_m$.  The multimers are
responsible for regulating gene expression and are allowed to bind to
the genome at an operator site.  Multimers are introduced in order to
get sufficient co-operativity in the binding isotherms to make a
genetic switch \cite{CA}, as described in more detail below.  The
state of the operator is represented by $\OO$, $\OO\TFA_n$,
\latin{etc}, depending on the state of binding of the multimers.  The
chemical reactions are
\begin{subequations}\label{chemeqs}
\begin{align}
&n\TFA \rightleftharpoons\TFA_n,\quad
m\TFB \rightleftharpoons\TFB_m,&&(\kf, \kb)\label{eqmulti}\\ 
&\OO+\TFA_n\rightleftharpoons\OO\TFA_n,\quad
\OO+\TFB_m\rightleftharpoons\OO\TFB_m,&&(\kon, \koff)\label{eqtra}\\ 
&\OO\TFA_n+\TFB_m\;|\;\OO\TFB_m+\TFA_n
\rightleftharpoons\OO\TFA_n\TFB_m,&&(\kon, \koff)\label{eqtrd}\\
&\OO\;|\;\OO\TFA_n\hookrightarrow\TFA,\quad
\OO\;|\;\OO\TFB_m\hookrightarrow\TFB,&&(\kA), (\kB)\label{eqexpr}\\
&\TFA\;|\;\TFB\to\nothing,&&(\muA), (\muB)\label{eqdegr}
\end{align}
\end{subequations}
Here we have adopted a condensed notation in which `$|$' indicates
alternative sets of reactants and `$\hookrightarrow$' indicates that
the reactants are not destroyed by the reaction.  For example the
first of Eqs.~\eqref{eqexpr}, $\OO\;|\;\OO\TFA_n\hookrightarrow\TFA$,
represents two reactions $\OO\to\OO+\TFA$ and
$\OO\TFA_n\to\OO\TFA_n+\TFA$.

The reactions in Eqs.~\eqref{chemeqs} account for, respectively, the
formation of multimers, the binding of TF multimers to the operator
(Eqs.~\eqref{eqtra} and \eqref{eqtrd}), the expression of TF monomers,
and the degradation of TF monomers.  Repression of gene expression is
implicit in Eqs.~\eqref{eqexpr}, thus $\TFA$ is expressed if and only
if $\TFB_m$ is not bound, \latin{etc}.  Reaction rates are as
indicated, and we define equilibrium constants for multimerisation,
$\Kd=\kf/\kb$, and operator binding, $\Kb=\kon/\koff$.

Whilst detailed and biologically faithful models can be constructed as
has been done for the \lphage\ switch \cite{ARMcA, ABJS}, the above
model is intentionally as simple as possible.  We believe that such an
approach is as important as detailed biological modelling in
elucidating the basic physical principles behind switches.  Thus, for
example, we have condensed the details of transcription and
translation into a single reaction step in Eqs.~\eqref{eqexpr},
governed by rate coefficients $\kA$ and $\kB$.  In a later section we
explore the possibility of refining the model at this point.

It should be noted, however, that the design of the network has to
obey certain constraints. In particular, the TFs must bind
co-operatively to the DNA in order to make a working switch
\cite{CA}. In the present model, co-operativity is introduced through
the binding of TF multimers rather than monomers.

\begin{table}
\caption{Distinct possibilities for the subsets of operator states for
our switch model: `0' indicates the state is disallowed, `1' indicates
it is allowed.\label{tab:poss}
}
\begin{ruledtabular}
\begin{tabular}{lcccc}
operator states  & $\OO$ & $\OO\TFA_n$ & $\OO\TFB_m$ &
$\OO\TFA_n\TFB_m$ \\
case & & $\alpha$ & $\beta$ & $\gamma$ \\
\hline
general                & \tick  & \tick  & \tick  & \tick  \\
exclusive              & \tick  & \tick  & \tick  & \cross \\
partially co-operative & \tick  & \tick  & \cross & \tick  \\
totally co-operative   & \tick  & \cross & \cross & \tick  \\
\end{tabular}
\end{ruledtabular}
\end{table}

In our model the operator is in one of four states $\{\OO, \OO\TFA_n,
\OO\TFB_m, \OO\TFA_n\TFB_m\}$.  We now include the effect of
interference between the upstream regulatory domains by disallowing
some of these states.  This is in the spirit of simplicity, strictly
speaking the effect is to modify the probabilities of the states.
Since the empty operator state is always a possibility and both
$\TFA_n$ and $\TFB_m$ should be allowed to bind otherwise they would
not be TFs, it turns out that there are only five possibilities, two
of which are related by symmetry.  The four distinct cases are shown
in Table \ref{tab:poss}, and are implemented by excluding some of the
reactions in Eqs.~\eqref{eqtra} and \eqref{eqtrd}.  For example, the
exclusive switch is obtained by discarding the reactions in
Eqs.~\eqref{eqtrd} thereby removing the state $\OO\TFA_n\TFB_m$.

\section{Mean-field analysis}\label{sec:mft}
We first analyse the behaviour of Eqs.~\eqref{chemeqs} using chemical
rate equations. This plays the role of a mean field theory for this
problem since chemical rate equations describe the temporal evolution
of the mean concentrations of molecules.  Switching behaviour
corresponds to the appearance of two distinct stable fixed points
(attractors) in the space of concentration variables.  For the general
switch, the problem has been analysed in detail by Cherry and Adler
\cite{CA}.  For the exclusive switch, a specific example has been studied
by Kepler and Elston \cite{KE}.  Our approach is a generalisation of
the analysis of Cherry and Adler.

Let us consider what happens for $\TFA$.  Write $\num{\TFA}$ for the
number of $\TFA$ monomers in the cell \latin{etc}, and let the cell
volume be $\Vcell$.  In a steady state, the multimerisation reaction
$n\TFA\rightleftharpoons\TFA_n$ is in equilibrium and
\begin{equation}
\Bigl(\frac{\num{\TFA_n}}{\Vcell}\Bigr)
=\Kd\,\Bigl(\frac{\num{\TFA}}{\Vcell}\Bigr)^n.
\end{equation}
Similarly the binding reaction $\OO+\TFA_n\rightleftharpoons\OO\TFA_n$
(if it is allowed) is in equilibrium and
\begin{equation}
\Bigl(\frac{\num{\OO\TFA_n}}{\Vcell}\Bigr)
=\Kb\,\Bigl(\frac{\num{\OO}}{\Vcell}\Bigr)\,
\Bigl(\frac{\num{\TFA_n}}{\Vcell}\Bigl).
\end{equation}
Now, $\num{\OO}$ is the probability that the operator is in state $\OO$
times the number of copies of the genome in the cell (usually assumed
to be one), \latin{etc}.  Therefore
\begin{equation}
\frac{\Prob(\OO\TFA_n)}{\Prob(\OO)}
=\frac{\Kb\Kd}{\Vcell^n}\,[\num{\TFA}]^n\equiv x^n
\label{bindeq}
\end{equation}
where we have introduced $x$ as a reduced concentration variable,
equal to $(1/\Vcell)(\Kb\Kd)^{1/n}$ times the number of monomers of
$\TFA$ in the cell.  Similarly we introduce a reduced concentration
variable $y$ for the number of monomers of $\TFB$.

>From the totality of binding equilibria, we surmise that the
probabilities of the operator states $\{\OO, \OO\TFA_n, \OO\TFB_m,
\OO\TFA_n\TFB_m\}$ are in the ratio $1:\alpha x^n:\beta y^m:\gamma
x^ny^m$ where we have covered off all the switch construction
possibilities by introducing a set of coefficients
$(\alpha,\beta,\gamma)$ which take the values zero or one according to
Table~\ref{tab:poss}.  The probability of the operator being in a state
where $\TFA$ is expressed is therefore
\begin{equation}
f(x,y) = \displaystyle{\frac{1+\alpha x^n}
{1+\alpha x^n+\beta y^m+\gamma x^ny^m}},\label{fdefeq}
\end{equation}
and of being in a state where $\TFB$ is expressed is
\begin{equation}
g(x,y) = \displaystyle{\frac{1+\beta y^n}
{1+\alpha x^n+\beta y^m+\gamma x^ny^m}}.\label{gdefeq}
\end{equation}
For $\TFA$, expression occurs at a rate $\kA$ and degradation at a
rate $\muA\times\num{\TFA}$.  It is convenient to
define a reduced degradation rate
\begin{equation}
\tmuA=\frac{\Vcell}{(\Kb\Kd)^{1/n}}\,\frac{\muA}{\kA}.
\end{equation}
If there is more than one copy of the genome in the cell, we should
additionally divide by the number of copies of the genome.  A similar
reduced degradation rate $\tmuB$ is defined for $\TFB$.

At a fixed point (a steady state), the rate of expression is equal to
the rate of degradation (for example $\kA f=\muA\num{\TFA}$), for both
TFs.  Using the reduced concentrations and degradation rates defined
above, the steady states are defined by 
\begin{equation}
f(x,y) = \tmuA x,\quad
g(x,y) = \tmuB y.\label{mfeqs}
\end{equation}
We see that the steady states only depend on $\tmuA$ and $\tmuB$.

The simplest way to proceed now is to consider the fixed points for
the dynamical system
\begin{equation}
\dot x = f(x,y) - \tmuA x,\quad
\dot y = g(x,y) - \tmuB y.\label{mf3eqs}
\end{equation}
This is only an approximation to the full dynamics, since it assumes
that TF binding is always in equilibrium.  However, Cherry and Adler
demonstrate under a mild restriction that the fixed points for
Eqs.~\eqref{mf3eqs} correspond to the possible fixed points of the
full system.  The mild restriction is that, under the full dynamics,
if the concentration of one of the TFs is held fixed, the
concentration of the other TF should settle on a unique value.  This
holds for our switch problem.

Switching behaviour corresponds to the existence of two stable fixed
points for Eqs.~\eqref{mf3eqs}, separated by a third fixed point which
is unstable in one direction, like a transition state.  Thus a test
for switching behaviour is whether the dynamical system in
Eqs.~\eqref{mf3eqs} admits an unstable fixed point which is unstable
in one direction.  From dynamical systems theory, this can be
determined by considering the determinant of the stability matrix
\begin{equation}
\mathbf{M} = 
\left(\begin{array}{cc}
f_x - \tmuA & f_y        \\
g_x        & g_y - \tmuB \\
\end{array}\right)\label{mdefeq}
\end{equation}
where $f_x=\partial f/\partial x$, \latin{etc}.  The required test is
that $\mathrm{det}\,\mathbf{M} < 0$ at the fixed point in question.
At a fixed point, one can eliminate $\tmuA$ and $\tmuB$ between
Eqs.~\eqref{mfeqs} and \eqref{mdefeq}.  We find that
$\mathrm{det}\,\mathbf{M} = (1/xy)\times D$ where we have introduced
the discriminant function
\begin{equation}
D(x,y) = (f-x f_x)(g-y g_y)-(x g_x)(y g_y).
\end{equation}
The sign of $D$ mirrors the sign of $\mathrm{det}\,\mathbf{M}$ so that
if $D<0$ at a fixed point, that fixed point is unstable in one
direction.  Moreover, for any point in the $(x,y)$ plane, $D$ has a
definite value independent of the values of $\tmuA$ and $\tmuB$.  Thus
in the plane of reduced concentration variables, the nature of a fixed
point is determined solely by its position.  In particular, the region
$D<0$ contains all the unstable fixed points of the kind desired, and
only those kind of fixed points.  This region is bounded by the line
$D=0$.

\begin{figure}
\begin{center}
\includegraphics{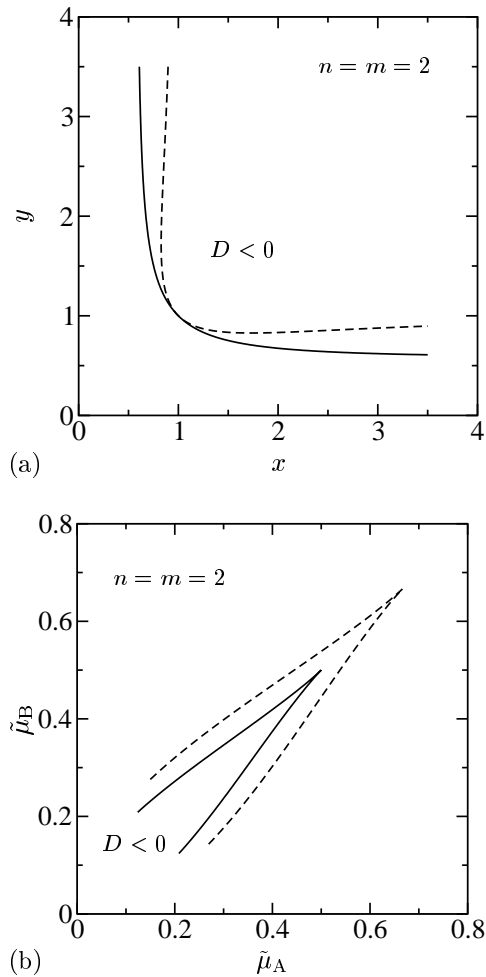}
\end{center}
\caption{In mean field theory, switching behaviour is confined to a
region in the $(x,y)$ plane in (a), or equivalently to a wedge in the
$(\tmuA,\tmuB)$ plane in (b).  Results are shown for dimerising
general (solid line) and exclusive (dashed line)
switches. Note that the bistable region is larger for the exclusive
switch than for the general switch.\label{fig:mftheory}
}
\end{figure}

Normally one regards the parameters $\kA$, \latin{etc}, as given, and
one has to solve Eqs.~\eqref{mfeqs} for the position of any fixed
points.  However, if one knows a fixed point $(x,y)$ in the plane of
reduced concentration variables, the corresponding reduced parameters
are given by $\tmuA=f/x$ and $\tmuB=g/y$, from Eqs.~\eqref{mfeqs}.
The region $D<0$ in the $(x,y)$ plane is therefore mapped to a region
(a wedge) in the $(\tmuA, \tmuB)$ plane.  Fig.~\ref{fig:mftheory}
shows two examples.  Note that the mapping from $(x,y)$ to $(\tmuA,
\tmuB)$ is triple-valued within the wedge, since for parameters where
switching behaviour occurs there are two stable fixed points in
addition to the unstable fixed point.

An interesting analogy with a fluid-fluid demixing transition in a
binary liquid mixture can be made at this point since both are
examples of a cusp catastrophe \cite{catastrophe}.  In this analogy,
$\tmuA^{-1}$ and $\tmuB^{-1}$ correspond to the chemical potentials of
the two components and the wedge $D<0$ to the region of spinodal
instability.  The cusp of the wedge in Fig.~\ref{fig:mftheory}(b)
would correspond to the critical point.  

We now apply the above analysis to the switch models of the preceding
section.  Firstly, for $n=m=1$, one can prove that $D>0$ 
for all four kinds of switch in Table~\ref{tab:poss}.  This
confirms the result of Cherry and Adler, namely that some form of
co-operativity is required to make a working switch.

Secondly, the discriminant for the totally co-operative case turns out
to simplify to $D=(1+z)[1+(m+n)z]$ where $z=x^n y^m$.  This is
positive for all values of $n$ and $m$.  Thus switching
behaviour cannot occur for the totally co-operative switch.

For the remaining cases, we have analysed in detail the situation for
\emph{dimerising} switches, with $n=m=2$.  The details of this
analysis are given in Appendix \ref{sec:app1}.
Fig.~\ref{fig:mftheory} shows the regions in the both the $(x, y)$ and
$(\tmuA,\tmuB)$ planes where switching behaviour occurs, for the
general and exclusive switches.  Clearly there is a more extensive
region of switching behaviour in the $(\tmuA,\tmuB)$ plane for the
exclusive case compared to the general case.  Switching behaviour is
strongly suppressed for the partially co-operative cases; for example
if $\OO\TFB_2$ is disallowed, switching occurs only if $\tmuA\alt0.1$
and $\tmuB\alt0.01$ (lines not shown in Fig.~\ref{fig:mftheory}).
Thus we conclude that, at least in mean field theory, restricting the
set of operator states can have a marked effect on the possibility to
form a genetic switch.

\section{Stochastic simulations}
\label{sec:simulations}
Whilst mean field theory indicates the regions in parameter space
where switching might occur, it has nothing to say about the switch
stability and the effects of number fluctuations in systems with a
finite size.  To address these problems, we therefore turn to a
stochastic simulation of the chemical reactions in
Eqs.~\eqref{chemeqs}.  This is done via the `Gillespie' algorithm
\cite{Gillespie}, which is a kinetic Monte Carlo scheme \cite{Bortz75}
to generate trajectories in the space of concentration variables
appropriate to the chemical master equation.

\begin{table}
\caption{Rate coefficients and equilibrium constants for a
representative set of the reactions which define the baseline
model.  Here $\Vcell$ is the cell volume and $k$ is used as a unit of
(inverse) time.\label{tab:rates}
}
\begin{ruledtabular}
\begin{tabular}{llll}
reaction & \multicolumn{2}{l}{baseline} \\
\hline
$\OO\to\OO+\TFA$ & $k$ & & \\
$\TFA\to\nothing$ & $\mu$ & $(0.2\hbox{--}0.8)\,k$ & \\
\hline
$2\TFA\to\TFA_2$ & $\kf$ & $5\kVcell$ & \\
$\TFA_2\to2\TFA$ & $\kb$ & $5k$ & $\Kd=\kf/\kb=\Vcell$ \\
\hline
$\OO+\TFA_2\to\OO\TFA_2$ & $\kon$ & $5\kVcell$ & \\
$\OO\TFA_2\to\OO+\TFA_2$ & $\koff$ & $k$ & $\Kb=\kon/\koff=5\Vcell$ \\
\end{tabular}
\end{ruledtabular}
\end{table}

Since these simulations are quite intensive, we focus again on the
dimerising general and exclusive switches, and on the symmetry line
$\muA=\muB=\mu$ and $\kA=\kB=k$.  We additionally have to supply
values for all the rate coefficients in Eqs.~\eqref{chemeqs}.  To do
this, we regard $k\approx0.1$--$1\,\mathrm{s}^{-1}$ (the expression
rate) as a unit of (inverse) time, and $\mu$ (the degradation rate) as
the main control variable.  As is the case with biological systems,
real rate coefficients vary over quite a range of values.  The
baseline parameter set is biologically motivated, with expression
being a slow step and binding equilibrium biased in favour of bound
states.  Rate coefficients for our baseline model for a representative
set of reactions are given in Table~\ref{tab:rates}.  For comparison,
literature values for \lphage\ can be found in Arkin \latin{et al}
\cite{ARMcA}, in Aurell \latin{et al} \cite{AS,ABJS}, and in Bundschuh
\latin{et al} \cite{BHJ}.  The rate coefficients for bimolecular
reactions, such as the forward dimerisation and forward binding
reactions, have units of volume divided by time.  Since our
concentrations are expressed as the number of molecules in the cell
volume, the natural units for these rate coefficients are $\kVcell$
where $\Vcell\approx 2\,\mu\mathrm{m}^3$ ($1/\Vcell\approx
1\,\mathrm{nM}$) is the cell volume.  For the baseline parameters, the
mean-field theory predicts bistability for $\mu/k<\surd{5}/2=1.12$ for
the general switch and $\mu/k<2\surd{5}/3=1.49$ for the exclusive
switch. The implementation of the Gillespie algorithm with
concentrations expressed as the number of molecules in the cell is
straightforward.

In the next section, we will first focus on the steady-state
properties of both switches. In the subsequent section, we will
address the dynamics of switching.

\subsection{`Free-energy' landscapes}

We monitor the total number of molecules of $\TFA$ and $\TFB$ in the
cell, $\NA$ and $\NB$.  These include the monomers, the free dimers,
and the bound dimers (see for example Eq.~\eqref{naeq}).  This is
motivated primarily by the observation by Bialek that the switch
lifetime is likely to be an exponential function of the number of
molecules involved in the switching process \cite{Bialek}.  Moreover,
$\NA$ and $\NB$ are the relevant slow variables in the sense of
Bundschuh \latin{et al} \cite{BHJ}.

\begin{figure}
\begin{center}
\includegraphics{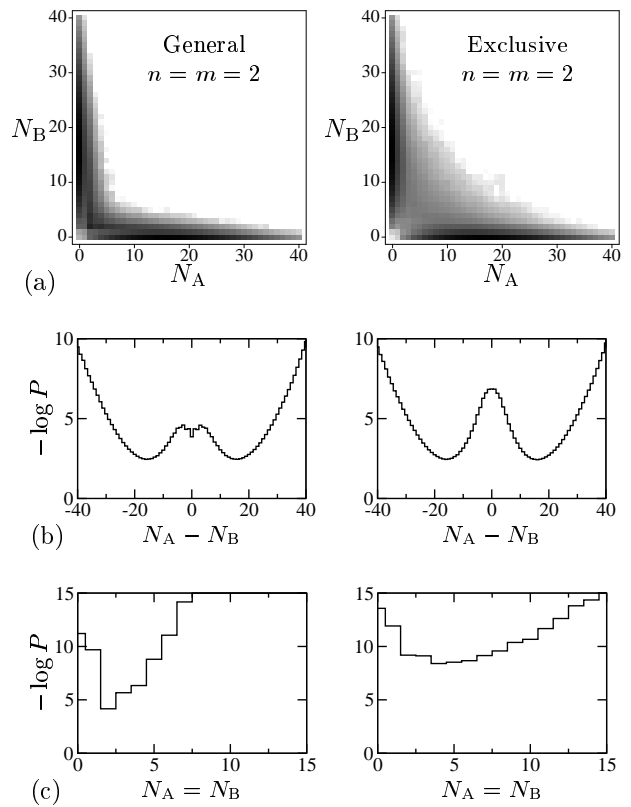}
\end{center}
\caption{(a) Probabilities in $(\NA,\NB)$ plane constructed from
trajectories of total duration $kt\approx8\times10^6$, for general and
exclusive switches, at $\mu=0.4\,k$.  Greyscale is logarithmic from
$P\le10^{-7}$ (white) to $P\ge0.04$ (black).  (b) Probabilities
collapsed onto the $\NA-\NB$ line, plotted as a dimensionless `free
energy' $-\log P$; the maximum at $\NA=\NB$ corresponds to the
mean-field transition state.  (c) Probabilities along the diagonal
line $\NA=\NB$; the minimum in this section now corresponds to the
mean-field transition state.  It is seen that the saddle-point of the
exclusive switch is higher than that of the general switch. See also
Table~\ref{tab:features} for locations of the features in
$P(\NA,\NB)$ and the barrier heights.\label{fig:densmap}
}
\end{figure}

To demonstrate the appearance of switching, we can construct the
probability distribution $P(\NA,\NB)$ for states in the $(\NA,\NB)$
plane.  This is done by generating a sufficiently long Gillespie
trajectory to obtain good coverage of all the interesting features of
the probability distribution \cite{note4}.  Fig.~\ref{fig:densmap}
shows the main features for $\mu=0.4\,k$, for both the dimerising
general and exclusive switches with this direct method.

\begin{table}
\caption{Locations of maxima and transition states for $P(\NA, \NB)$
for dimerising general (DGS) and exclusive (DES) switches at
$\mu=0.4\,k$ (Fig.~\ref{fig:densmap}), compared to fixed points from
mean field theory (MFT, section \ref{sec:mft} and appendix
\ref{sec:app1}).\label{tab:features}
}
\begin{ruledtabular}
\begin{tabular}{llllc}
 & & \multicolumn{2}{c}{Maximum} & Transition state \\
 & & $\NA$ & $\NB$ & $\NA=\NB$ \\
\hline
      DGS   & Gillespie & $15.0\pm0.5$ & $0$     & $2\pm1$ \\
$(\gamma=1)$ & MFT   & $16.91$        & $0.09$ & 5.74 \\
\hline
      DES   & Gillespie & $15.0\pm0.5$ & $0$     & $4\pm1$ \\
$(\gamma=0)$ & MFT   & $16.04$        & $0.16$ & 3.15 \\
\end{tabular}
\end{ruledtabular}
\end{table}

We see that switching appears as a double maximum in the probability
distribution, and there is a transition state at a low number of
molecules for both TFs; we {\em assume} here that the transition state
corresponds to the saddle point in the `free-energy landscape'
$-\log\left [P(N_A,N_B)\right]$. Table~\ref{tab:features} contains the
locations of the probability maxima and the transition state,
comparing the Gillespie results with the mean-field theory fixed point
solutions of the preceeding section.  The maxima are in quite good
agreement with the mean-field fixed points, although there is some
difference in the location of the saddle-point.  The first column in
Table \ref{tab:barrier} shows the probability of reaching states with
$\NA=\NB$ determined from Fig~\ref{fig:densmap}.  This probability is
an order of magnitude lower for the exclusive switch than for the
general switch, which indicates that the exclusive switch is more
stable than the general switch.

\begin{figure}[b]
\begin{center}
\includegraphics{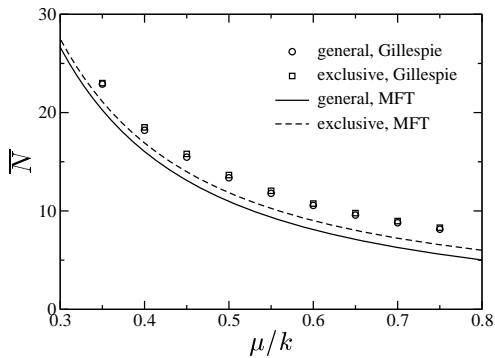}
\end{center}
\caption{Mean copy number of most-expressed TF, as a function of the
degradation rate, for both kinds of switch and baseline
parameters. `Gillespie' refers to the results of the stochastic
simulations and `MFT' refers to those of the mean-field analysis
described in section~\ref{sec:mft}.\label{fig:numvsk}
}
\end{figure}

Fig.~\ref{fig:numvsk} shows the mean number of molecules of the
most-expressed transcription factor as a function of $\mu/k$ in the
switching regime, for the two kinds of switch.  We define this
\latin{via} the time average
\begin{equation}
\Nmean = \overline{\mathrm{max}(\NA,\NB)}.
\end{equation}
Since the system spends most of the time near a stable state, this
average is dominated by the number of molecules in the stable
states. As such, the time average can be compared to the mean field
stable fixed point.  Further, in the stable steady states both
switches should behave very similarly
(see Eq.~\ref{chemeqs}). Fig.~\ref{fig:numvsk} shows that while the
average number of molecules $\Nmean$ is indeed nearly the same for the two
switches, it also lies systematically a little above the mean field
theory prediction.

\subsection{Rate constants}
\label{sec:Rates}
\begin{figure}
\begin{center}
\includegraphics{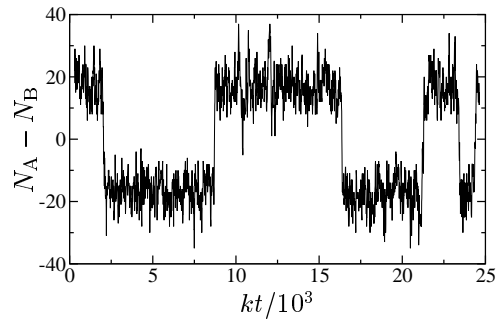}
\end{center}
\caption{Typical time series for $\NA-\NB$.  Data is for the
dimerising exclusive switch at $\mu=0.4\,k$.\label{fig:signal}
}
\end{figure}

We now turn to the switch {\em dynamics}.  Fig.~\ref{fig:signal} shows
typical time series for $\NA-\NB$ for the dimerising exclusive
switch at $\mu=0.4\,k$.  The appearance of the two switch states
where one of the TFs is strongly repressed compared to the other one
is clearly seen.  These correspond to the probability maxima of
Fig.~\ref{fig:densmap}(a).  A switching event occurs when the roles of
the two TFs flip spontaneously.

In order to elucidate the stability of the switches, we make a
connection with the reactive flux method that has been pioneered by
Bennett \cite{Bennett77} and Chandler \cite{Chandler78} and that is
now widely used in the field of soft-condensed matter physics
\cite{Chandlerbook}. We first define an order parameter, $q$, that
serves as our reaction coordinate and measures the progress of
flipping the switch from one state to the other. In what follows, we
will take $q= N_A - N_B$. Furthermore, we define two characteristic
functions that indicate in which state the system is in:
\begin{eqnarray}
h_A (q) &=& 1, \, \,  q < q^*\\
&=& 0, \, \,q \geq q^*\\
h_B(q) &=& 0, \, \, q < q^*\\
&=& 1, \, \, q \geq q^*.
\end{eqnarray}
It is natural to take for $q^*$ the value that corresponds to the top
of the `barrier' that separates the two stable steady sates. We have
thus chosen $q^*=0$ for both switches (see Fig.~\ref{fig:densmap}).

We now assume that we can make a separation of time scales
\cite{Chandlerbook}. In particular, we assume that there exists a time
$t$ that is longer than the time $t_{\rm trans}$ it takes for
transient behavior to relax, but shorter than the characteristic time
$t_{\rm rxn}$ for making a transition from one state of the switch to
the other. If this assumption holds, then we can coarse-grain the
switch as a two-state system. If the flipping on the time scale
$t_{\rm rxn}$ is a Poisson process, then the relaxation of the switch
is given by the following expression
\begin{equation}
\label{eq:hahb}
\frac{\overline{h_A (0) h_B (t) }}{\overline{h_A}}
 =\overline{h_B} \left(1- \exp[-(k_{AB} + k_{BA}) t]\right).
\end{equation}
Here, the overbar denotes a time average; contributions to the average
are weighted according to the stationary distribution of states. The
expression on the left-hand side yields the probability that
the switch is in state $B$ at time $t$, given that initially it was  in
state $A$. If this time $t$ is longer than  $t_{\rm trans}$
and if there are no correlations between successive switching events
on this time scale, then this correlation function should be given by
the macroscopic expression on the right-hand side of the above
equation, where $k_{AB}$ and $k_{BA}$ are the rate constants for
flipping the switch in the forward and backward directions,
respectively. The above relation thus only holds for $t > t_{\rm trans}$.

It is instructive to take the time derivative in
Eq.~\ref{eq:hahb}. This yields
\begin{equation}
\frac{\overline{h_A (0) \dot{h}_B (t)}}{\overline{h_A}} = 
\overline{h_B} (k_{AB} + k_{BA}) \exp[-(k_{AB} + k_{BA})t].
\end{equation}
If we now consider times $t$ smaller than $t_{\rm rxn} = 1/(k_{AB} +
k_{BA})$, but larger than $t_{\rm trans}$, and exploit the detailed
balance condition $\overline{h_B}/\overline{h_A} = k_{AB}/k_{BA}$,
then we find
\begin{equation}
\frac{\overline{h_A (0) \dot{h}_B (t)}}{\overline{h_A}} = k_{AB}.
\end{equation}
This shows that the flux of trajectories from $A$ to $B$ should
exhibit a plateau for times $t_{\rm trans} < t <
1/(k_{AB}+k_{BA})$. Indeed, the macroscopic switching rate is
precisely given by the constant flux of trajectories in this regime.

In general, however, it is useful to define a time dependent rate constant:
\begin{eqnarray}
k_{AB}(t) &=& \frac{\overline{h_A (0) \dot{h}_B (t) }}{\overline{h_A}}\\
&=& \frac{\overline{ \dot{q}(0) \delta(q(0) - q^*)
\theta(q(t)-q^*) }}{\overline{ \theta(q^*-q(t))}}\\
&=&\frac{\overline{\delta
(q-q^*)}}{\overline{ \theta(q^*-q(t)) }}\frac{\overline{ \dot{q}(0) \delta(q(0)-q^*)
\theta(q(t)-q^*)}}{\overline{ \delta (q-q^*) }}\\
&=& P_{\rm 0}(q^*) \; \overline{ \dot{q}(0) \theta(q(t)-q^*)}^{*}\\
&=&P_{\rm 0}(q^*) \; R(t). \label{eq:PoR}
\end{eqnarray}
Here, $\theta$ is the Heaviside function. The overline with the
asterix $^*$ denotes an average over states at the top of the barrier.

It is seen that $k(t)=k_{AB}(t)$ is the product of two
contributions. The first is $P_{\rm o}(q^*)$, which is given by,
\begin{equation}
P_{\rm 0} (q^*) = \frac{P(q^*)}{\sum_{-\infty}^{q^*} \, \, P(q)}.
\end{equation}
Here $P(q)$ is the steady-state probability of finding the system with
a reaction coordinate of size $q$. Noting that if $q<q^*$ the system
is in the initial state, it is clear that $P_{\rm o}(q^*)$ is the
probability of finding the system at the top of the barrier divided by
the probability of finding it in the initial state. This quantity can
be directly obtained from Fig.~\ref{fig:densmap}, and is shown in
Table~\ref{tab:barrier}. 

The second contribution to $k(t)$ is $R(t)$, which gives the average
flux of trajectories at the top of the barrier. As explained in
\cite{Chandler78}, the {\em initial} rate $k(t\rightarrow 0^+)$ corresponds to
the approximation for the rate constant in the transition state theory
of chemical reactions:
\begin{equation}
k_{\rm TST} = k (t\rightarrow 0^+) = P_{\rm 0}(q^*) \; 
\overline{ \dot{q}(0)\theta(\dot{q})}^{*}.
\end{equation}
Transition-state theory assumes that all trajectories
initially heading from the top of the barrier towards the final state
will indeed end up in the final state and all trajectories initially
heading towards the initial state will end up in the initial
state. This assumption is only correct if no trajectories recross
the barrier. In the present case, recrossing turns out to be quite
significant and, as a result, $k(t)$ will be significantly lower than
$k_{\rm TST}$. It is conventional to express the reduction of $k(t)$
due to recrossings in terms of the transmission
coefficient $\kappa(t)$, defined as
\begin{equation}
\kappa(t) \equiv \frac{k(t)}{k_{\rm TST}} = \frac{R(t)}{R(0^+)}
\end{equation}
where $R(0^+)=\overline{ \dot{q}(0)\theta(\dot{q})}^{*}$.
We note that, just as $k_{AB}(t)$ exhibits a plateau value for $t_{\rm
trans} < t < t_{\rm rxn} = 1/(k_{AB} + k_{BA})$, $\kappa (t)$ and
$R(t)$ also reach a constant value on this time scale. We will simply
refer to them as the transmission coefficient $\kappa$ and kinetic
prefactor $R$, respectively.

\begin{figure}
\begin{center}
\includegraphics{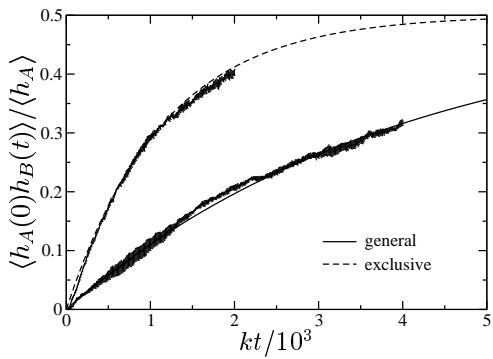}
\end{center}
\caption{The cross correlation between the indicator functions
$h_A=\theta(q_A-q(t))$ and $h_B=\theta(q(t)-q_B)$ allows an accurate
determination of the switch lifetime.  Results are shown for the
general and exclusive switches at $\mu=0.4\,k$, with $-q_A=q_B=5$.
The lines are fits to Eq.~\eqref{eq:hahb} in the form $\langle
h_A(0)h_B(t)\rangle/\langle h_A\rangle=(1-\exp[-2t/\tau])/2$.  Table
\ref{tab:barrier} contains the corresponding values of
$\tau$.\label{fig:hahb}
}
\end{figure}

For many rare event problems in soft-condensed matter, it is not
feasible to obtain the rate constant by calculating the correlation
function $\langle h_A(0) h_B(t) \rangle $ in a long, brute-force
simulation. In order to alleviate the rare event problem, a widely
used approach is to first calculate the free-energy barrier $P_o(q^*)$
using umbrella sampling \cite{Torrie74} and then compute the kinetic
prefactor $R(t)$ by shooting-off (molecular dynamics) trajectories
from the top of the barrier \cite{Bennett77,Chandler78}. However, the
umbrella sampling technique in its conventional form relies on an
energy-functional and is therefore only applicable to systems that
obey detailed balance. Biochemical networks are usually out of
equilibrium and consequently lack detailed balance. In Appendix B, we
show that the genetic toggle switches considered here also do not obey
detailed balance. We can therefore not use the Bennett-Chandler
approach \cite{Bennett77,Chandler78}. Indeed, we have resorted to
long, brute-force simulations in order to calculate the flipping rates
for the toggle switches. To be more precise \cite{note4}, we have
obtained the rate constant by fitting the correlation function
$\overline {h_A(0) h_B(t)}/\overline{h_A}$ to its macroscopic
expression as given by Eq.\ref{eq:hahb}; for the symmetric switches
considered here, $k_{AB} = k_{BA} = 1/\tau$, where $\tau$ is the life
time of the stable steady state. In practice, to reduce noise in the
correlation function, we excise a window of states around $q^*$ and
define $h_A=\theta(q_A-q(t))$ and $h_B=\theta(q(t)-q_B)$ where
$q_A<q^*$ and $q_B>q^*$.  Most of our calculations have been performed
for $-q_A=q_B=5$, but we find that, as expected, our results are
insensitive to the precise values chosen, provided that the system is well
into the switching regime where there is a good separation of time
scales.  Fig.~\ref{fig:hahb} shows typical correlation functions.

\begin{figure}[b]
\begin{center}
\includegraphics{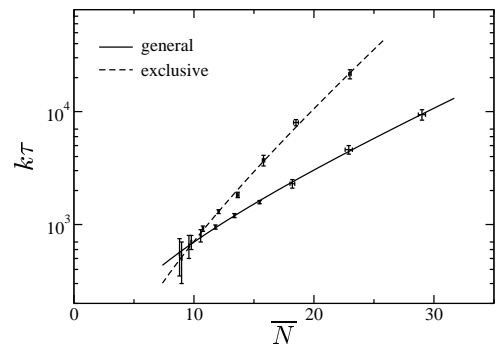}
\end{center}
\caption{Switch lifetime as a function of the mean number of the
most-expressed TF.  The exclusive switch becomes orders of magnitude
more stable than the general switch at high numbers of the expressed
TF.  The lines are fits to $\tau\sim\Nmean{}^{\alpha}\!\exp(b\Nmean)$
as discussed in the text.\label{fig:taunum}
}
\end{figure}

In Fig.~\ref{fig:taunum} we show the switch lifetime $\tau = 1/k_{AB}$
($=1/k_{BA}$ for the symmetric switches) as a
function of the mean value of the most-expressed TF for both kinds of
switches, as $\mu$ varies.  We see that $\tau$ grows extremely rapidly
with $\Nmean$, which is the basic reason why extremely stable switches
can be built with only a few hundred expressed proteins.  Our
simulations cover $10\alt\Nmean\alt30$, but if we extrapolate our
results to $\Nmean\approx100$, then $k\tau\approx10^7$ for the
general switch but $k\tau\approx10^{11}$ for the exclusive
switch.  In the latter case, this corresponds to lifetimes measured in
tens of years.  Such extremely long lifetimes have been reported for
\lphage\ \cite{ABJS}.  Equally important, Fig.~\ref{fig:taunum} is a
dramatic confirmation that the switch construction has a marked
influence on stability.  The exclusive switch lifetime grows much more
rapidly with the mean copy number than the general switch lifetime.

We find that $\tau$ grows sub-exponentially with $\Nmean$ and can be
fit to a functional form suggested by analysis of a related problem,
namely that of switching between broken-symmetry phases in a driven
diffusive model \cite{EFGM,GLEMSS}.  The fit is to
$\tau=A\Nmean{}^{\alpha}\!\exp(b\Nmean)$, where $A=50.4$,
$\alpha=0.72$, $b=0.097$ for the general switch, and $A=7.32$,
$\alpha=1.16$, $b=0.19$ for the exclusive switch.  This fit gives
quantitative support to Bialek's conjecture that the switch lifetime
grows exponentially with the number of molecules involved in the
switching process \cite{Bialek}, but additionally indicates that there
is a logarithmic correction in $\Nmean$.

\begin{table}
\caption{Switching kinetics.  The first column is the probability of
reaching the top of the barrier from Fig.~\ref{fig:densmap}, the
second column is the lifetime from Fig.~\ref{fig:hahb}, and the third
column is kinetic prefactor defined to be $R=1/(P_{\rm 0}(q^*)\tau)$.
The fourth column is the mean escape rate $R(0^+)$ to $\NA>\NB$ from
$\NA=\NB$ as determined from additional analysis of the simulations, and
the final column is the transmission coefficient $\kappa=R/R(0^+)$.  A
figure in brackets after a result is an estimate of the error in the
final digit of that result.\label{tab:barrier}
}
\begin{ruledtabular}
\begin{tabular}{lccccc}
 & $P_{\rm 0} (q^*)/10^{-3}$ & $k\tau/10^{3}$ 
 & $R/k$ & $R(0^+)/k$ & $\kappa$ \\
\hline
 DGS & $9(1)\phantom{.0}$   & $2.3(2)$  & $0.048(7)$
  & $0.24(1)$ & $0.20(3)$ \\
 DES & $0.92(1)$ & $8.0(5)$ & $0.14(1)\phantom{0}$ 
  & $0.98(4)$ & $0.14(1)$
\end{tabular}
\end{ruledtabular}
\end{table}

Table \ref{tab:barrier} shows the barrier heights $P_o({q^*})$
computed from the direct numerical simulations of $P(\NA,\NB)$. We can
now obtain the kinetic prefactor $R$ by dividing the rate constant
$k_{AB} = 1/\tau$ by $P_o(q^*)$: $R = k_{AB}/P_o(q^*)$ (see also
Eq. \ref{eq:PoR}). We find that the underlying barrier-crossing rate
$R \approx 0.05 - 0.15 k$ is a (small) fraction of $k$, which corresponds to
the time scale for gene expression (see Table
\ref{tab:rates}). In fact, for the general switch, the kinetic
prefactor $R$ is significantly lower than the slowest reaction in the
system, which is the degradation reaction with $\mu=0.2 - 0.8 k$ (see
also Table \ref{tab:rates}). Since $R$ can be interpreted as the rate
at which the barrier is crossed (i.e., the flux of trajectories at the
top of the barrier), it is clear that crossing the barrier typically
involves a large number of protein synthesis and degradation steps.

We have also directly computed from the Gillespie trajectories the
distribution of escape times from the dividing surface $\NA=\NB$
to states with $\NA\ne\NB$. We find that for both kinds of switches, the
escape time distribution is quite well approximated by a Poisson
distribution with a characteristic lifetime $\tauescape$.  Since the
switches are symmetric, this means we can define the escape rate to
one side (for example to $\NA>\NB$) to be $R(0^+) = 1/2\tauescape$.
Estimates for $R(0^+)$ from the Poisson-distribution fit are shown in
the fourth column in Table~\ref{tab:barrier}.  The ratio of the
barrier crossing rate to the one-sided escape rate is the transmission
coefficient, $\kappa=R/R(0^+)$ \cite{Chandlerbook}.  As
Table~\ref{tab:barrier} shows, the transmission coefficient for both
kinds of switch is 10--20\%.  This shows that the barrier crossing is
quite diffusive. In other words, a typical transition path between the
two stable regions crosses and re-crosses the top of the barrier
$\NA=\NB$ several times before committing to one of the stable basins.
This is in accord with our findings on the statistics of $q=0$
crossing times reported previously \cite{WtW04_1}. It is also
consistent with the observation that a typical transition path
involves many protein synthesis and degradation reactions.

\section{Beyond the baseline model}
We now consider the effects of varying parameters away from the
baseline set, and changing some other aspects such as introducing a
more detailed representation of transcription and translation.  Switch
stability is generally unaffected by variation of the kinetic rates
that do not alter the mean copy number of the TFs.  What does have an
effect, as we shall see, are phenomena which increase the `shot noise'
in expression.  These include the generation of multiple copies of the
TF from each mRNA transcript and stochastic fluctuations in the number
of copies of the TF generated per mRNA transcript.

In this section we give results for the dimerising exclusive switch.
We have repeated all the simulations for the dimerising general switch
and we find that the same trends are recovered.  We also find
that the exclusive switch is \emph{always} markedly more stable than
the corresponding general switch, in a manner represented by
Fig.~\ref{fig:taunum}.

\begin{figure}
\begin{center}
\includegraphics{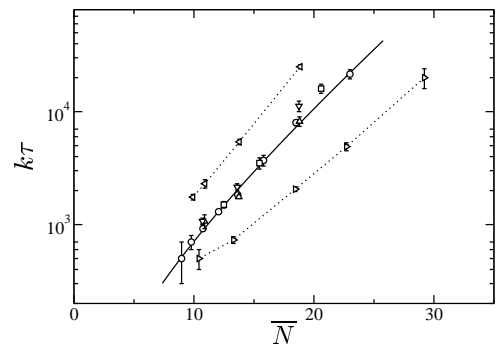}
\end{center}
\caption{Switch lifetime for the dimerising exclusive switch as some
of the baseline kinetic parameters are varied. Data is shown for
baseline model (circles; see Table~\ref{tab:rates}), with the number
of copies of the genome doubled and expression rate halved (squares),
with $\kf$ and $\kb$ doubled (up triangles), with $\kon$ and $\koff$
doubled (down triangles), with $\kon=10\kVcell$ and, as before, $\koff
= k$ (left triangles), and with $\kon=2\kVcell$ and, as before, $\koff
= k$ (right triangles); see Table~\ref{tab:rates} for the meaning of
the rate constants.  The solid line is the fit to the baseline data
from Fig.~\ref{fig:taunum}.\label{fig:varied} }
\end{figure}

\subsection{Varying parameters away from baseline}
Fig.~\ref{fig:varied} shows that the switch lifetime is very
insensitive to variations away from the baseline kinetic parameters
that do not affect the mean field steady states.  If, however, we vary
the kinetic parameters such that the mean field steady states are
changed, then we do see a systematic change in the switch stability.
For instance, the switch stability is enhanced if the binding
equilibrium is moved in favour of bound TF by increasing $\kon$ (the
rate of binding to DNA).  This is because the number of molecules of
each TF at the point where switching just starts decreases as $\Kb =
k_{\rm on} / k_{\rm off}$ increases (see Eq.~\eqref{nceq}), and thus at
fixed $\Nmean$ we are effectively moving deeper into the bistable
region. This leads to a more stable switch.

\subsection{Effect of messenger RNA}
\begin{figure}
\begin{center}
\includegraphics{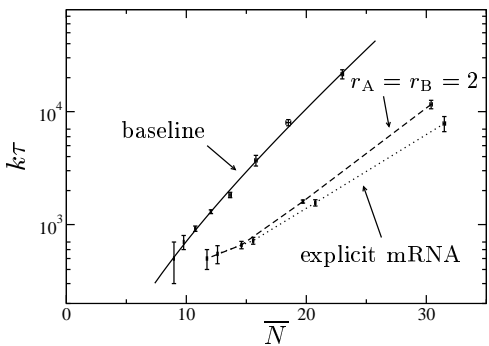}
\end{center}
\caption{Switch stability is degraded if multiple copies of the TF are
produced in each expression reaction, in this case $\rA=\rB=2$ in
Eq.~\eqref{eqrexpr}.  In a model which includes explicit mRNA
production (as described by Eqs.~\eqref{extreqs}), then even if on
average only {one} TF is produced per transcript, the stability is
similarly degraded by fluctuations about the
mean. `Baseline' refers to the baseline model, as described in
Table~\ref{tab:rates}. \label{fig:messenger}
}
\end{figure}

We have condensed the various steps in gene expression into a single
`expression reaction' in Eqs.~\eqref{eqexpr}. In reality, however, many steps
are involved.  We now analyse these steps in more detail.
Recall that the genetic information is first \emph{transcribed}
into messenger RNA (mRNA), then \emph{translated} into proteins by
ribosomes.  It is often the case that several copies of the protein
can be generated from one mRNA transcript.  In the simplest way, we
can capture this by replacing Eqs.~\eqref{eqexpr} by a version in
which $r$ copies of each transcription factor are generated from each
mRNA transcript, thus
\begin{equation}
\OO\;|\;\OO\TFA_n\hookrightarrow \rA\times\TFA,\quad
\OO\;|\;\OO\TFB_m\hookrightarrow \rB\times\TFB.\label{eqrexpr}
\end{equation}
It is easy to see that nothing changes in mean field theory provided
that $\rA\kA$ and $\rB\kB$ are used in place of $\kA$ and $\kB$.  We
have undertaken simulations for the case $\rA=\rB=2$ and
$\kA=\kB=k/2$, which in mean field theory is identical to our
baseline.  However, as Fig.~\ref{fig:messenger} shows, the switch
lifetime is decreased as a result of having TFs generated two at a
time.  The basic reason is that the `shot noise' in the expression
reactions has increased.

Even if on average one TF is produced from each mRNA transcript, noise
can arise from the fluctuations about this mean, as we now discuss.
To capture this, we explicitly include the generation and degradation
of mRNA for the two TFs in the list of chemical reactions.  The
expression reactions in Eq.~\eqref{eqexpr} are replaced by the
following reactions:
\begin{subequations}\label{extreqs}
\begin{align}
&\OO\;|\;\OO\TFA_n\hookrightarrow\MA\to\nothing,&&(\kdA), (\mudA)\\
&\OO\;|\;\OO\TFB_m\hookrightarrow\MB\to\nothing,&&(\kdB), (\mudB)\\
&\MA\hookrightarrow\TFA,\quad
\MB\hookrightarrow\TFB.&&(\ktA), (\ktB)\label{eqmess}
\end{align}
\end{subequations}
In these, $\MA$ and $\MB$ are the mRNA species, and the reactions
represent the generation and degradation of mRNA, and the translation
of mRNA into TFs.  We have introduced new rate coefficients for
the translation, transcription and mRNA degradation steps.

In a steady state in mean field theory, the rate of transcription
balances the rate of mRNA degradation and the mean number of mRNA
transcripts is $k'/\mu'$.  Each transcript produces TFs at a rate $k''$,
hence the overall rate of production of TF is $(k' / \mu') \times
k''$.  We conclude that the equivalent expression rate is
\begin{equation}
k=\frac{k'k''}{\mu'}.\label{rnaeq}
\end{equation}
Going beyond mean field theory, the statistics of this basic model
for mRNA translation can be solved exactly \cite{McAA,TvO}.
Translation occurs at a rate $k''$ and degradation at a rate $\mu'$,
hence the probability per unit time of either a translation or a
degradation event taking place is $k'' + \mu'$.  Given that an event
has taken place, the probability that it was a translation event is $p
= k'' / (k'' + \mu')$ and the probability that it was a degradation
event is $q = 1 - p = \mu' / (k'' + \mu')$.  Therefore, the probability
that $r$ copies of the protein are generated before the mRNA
transcript is degraded is $P_r = q p^r$.  From this, all the
statistics can be calculated.  For instance, the mean number of TFs
produced per transcript is $\langle r\rangle = \sum_{r=0}^\infty r P_r
= p / q = k'' / \mu'$, which agrees with the expectation from
Eq.~\eqref{rnaeq}.  Also $\langle r^2\rangle=p / q^2$, so
that the standard deviation divided by the mean is $\sqrt{q/p}=\langle
r\rangle^{-1/2}$.  Thus there is considerable noise due to
fluctuations in the number of proteins generated per transcript, and
this noise is largest when the mean number of TFs produced per
transcript is small.

To see the effect of this noise we have implemented the additional
reactions in Eqs.~\eqref{extreqs} in our reaction schemes and again
determined the switch lifetimes by direct simulation.
Fig.~\ref{fig:messenger} shows the results for $\ktA = \ktB = 5\kdA =
5\kdB = \mudA = \mudB = 5k$.  For these parameter values $\ktA\kdA /
\mudB = \ktB\kdB / \mudB = 1$ and, according to Eq.~\eqref{rnaeq}, such
a system should be identical to our baseline.  Moreover since
$\ktA/\mudA = \ktB/\mudB = 1$, there is on average one TF produced
per mRNA transcript, so the `shot noise' associated with the mean
number of TFs produced per transcript is the same as in the baseline
model.  However, the simulations clearly demonstrate that the
additional noise due to fluctuations about this mean reduces the
switch lifetime considerably.

\subsection{Expression as a multistep process}
As a second refinement to the model for gene transcription and
translation, we now consider gene expression as a multistep process.
We thus replace the single reaction
$\OO\to\OO+\TFA$ by a sequence of reaction steps
\begin{equation}
\OO\to\OO_1\to\OO_2\to\OO_3\to\dots\to\OO_n\to\OO+\TFA.\label{activeq}
\end{equation}
If there is only one intermediate stage, $\OO_1$ say, this can be used
to model the formation of an `open complex' \cite{trpbook}.  If there
are multiple intermediate stages, this could represent the individual
steps of the RNA polymerase that walks along the DNA, or those of the
ribosomes that walk along the mRNA.  In some of the more detailed
models of gene expression, all these intermediate steps are captured
in detail \cite{ARMcA}.

We can make a connection with the baseline model by computing the
waiting time for the lumped reaction $\OO\to\OO+\TFA$.  Since the
waiting times for the individual steps in Eq.~\eqref{activeq} are
independent statistical quantities, the waiting time for the whole
sequence is the sum of the waiting times for the individual steps.  In
terms of reaction rates, $1/k=\sum 1/k_i$, where $k$ is the rate of
the lumped reaction, and the $k_i$ are the rates of the intermediate
steps.

It follows that the waiting time distribution for the lumped reaction
is not a Poisson distribution.  Indeed the central limit theorem
indicates that the lumped reaction will tend to have a Gaussian
distribution of waiting times, effectively converging on a
$\delta$-function for a very large number of steps.  Thus the
approximation which replaces Eq.~\eqref{activeq} by $\OO\to\OO+\TFA$
amounts to replacing the true non-Poisson distribution of waiting
times by a Poisson distribution with the same mean.

We have tested the consequences of this assumption for two cases.  In
the first test, we have inserted a single intermediate stage in
Eq.~\eqref{activeq} to represent the formation of the open complex.
The rates of the two steps were chosen to be $(5/4)k$ and $5k$, so
that formation of the open complex is the slow step.  With this
choice, the rate for the lumped reaction is the same as the baseline
model.  In the second test, we have inserted four intermediate stages
in the reaction scheme, so that there are five intermediate reaction
steps between $\OO$ and $\OO+\TFA$, to model for example the
progressive stages of transcription.  We have chosen the rates of
these intermediate steps all equal to $5k$ so again the effective rate
for the lumped reaction is the same as the baseline model.  However,
the sum of the variances for the individual steps is now five times
smaller than the variance in the waiting time for the lumped reaction.

\begin{figure}
\begin{center}
\includegraphics{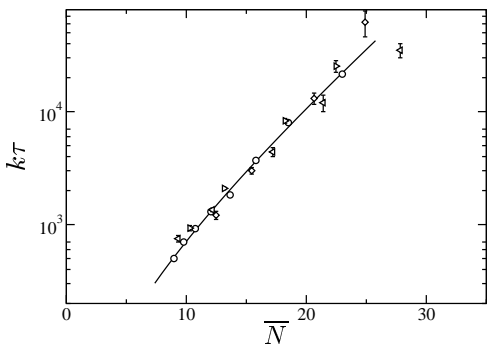}
\end{center}
\caption{Compared to the baseline model (circles; see Table~\ref{tab:rates}), there is
practically no effect on switch stability when one activation step is
included (right triangles) or several are included (left triangles) as
in Eq.~\eqref{activeq} (see main text for details).  There is also
little change if the operator is `split' as in Fig.~\ref{fig:split}
and Eqs.~\eqref{spliteqs} (diamonds).  The line is the fit to the
baseline data from Fig.~\ref{fig:taunum}.\label{fig:activesplit}
}
\end{figure}

Fig.~\ref{fig:activesplit} shows data for the dimerising exclusive
switch with these modifications.  It shows that the switch stability
is practically unaffected by the inclusion of the additional reaction
steps.  These results suggest that the precise waiting time statistics
for these multistep reactions are less important for the switch
stability than the statistics of the number fluctuations (as studied
in considerable detail in the preceeding section).  This is perhaps
not so surprising, since the activation process to flip the switch
must proceed by multiple coincidences of the `right' expression or
degradation events (see also discussion on kinetic prefactor $R$ in
section ~\ref{sec:Rates}), and so the detailed waiting time statistics
of the individual reaction steps are unimportant.

\subsection{Can the operator be split?}
\begin{figure}
\begin{center}
\includegraphics{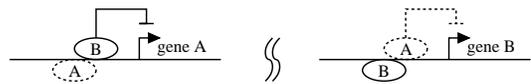}
\end{center}
\caption{A `split' exclusive switch can be built out of distinct
operators provided each TF blocks the other TF from
binding, at each operator site.\label{fig:split}
}
\end{figure}

The basic premise of the exclusive switch is that the binding of one
TF excludes the binding of the other TF.  Whilst it is natural to
think of this in terms of a diverging pair of genes on opposite
strands of the DNA as shown in Fig.~\ref{fig:switch}(b), it is also
possible to achieve the same effect by having each TF prevent the
binding of the other TF at separated operators on the DNA. An
exclusive switch with such a `split' arrangement is shown in
Fig.~\ref{fig:split}.  This arrangement is interesting because for
each gene, one TF regulates expression by acting as a repressor, but
the other TF acts only indirectly by preventing binding of the
repressor TF.

We can set up a system of chemical reactions to model the split
operator case as follows.  Suppose that $\OA$ and $\OB$ denote the
operator sites for the genes which code for $\TFA$ and $\TFB$,
respectively.  Then the set of reactions is
\begin{subequations}\label{spliteqs}
\begin{align}
&\OA+\TFA_n\rightleftharpoons\OA\TFA_n,\quad
\OA+\TFB_m\rightleftharpoons\OA\TFB_m,\label{spa1a}\\ 
&\OB+\TFA_n\rightleftharpoons\OB\TFA_n,\quad
\OB+\TFB_m\rightleftharpoons\OB\TFB_m,\label{spa1b}\\ 
&\OA\;|\;\OA\TFA_n\hookrightarrow\TFA,\quad
\OB\;|\;\OB\TFB_m\hookrightarrow\TFB,\label{spexpr}
\end{align}
\end{subequations}
plus the multimerisation and degradation reactions from
Eqs.~\eqref{chemeqs}.  We assume that the rates for all binding and
unbinding reactions are $\kon$ and $\koff$, and those for expression of
$\TFA$ and $\TFB$ are $\kA$, and $\kB$, respectively.  This set of
reactions describes the split analogue of the exclusive switch.  In
mean field theory, this model is identical to the standard case.  We
have simulated this model and indeed find that there is little
alteration to the switch stability, as shown in
Fig.~\ref{fig:activesplit}.

We conclude that it is the exclusive nature of the TF binding which
makes the exclusive switch markedly more stable than the general case.
This can be achieved by overlapping the operator sites of genes that
are arranged in a divergent manner and (thus) on opposite strands of
the DNA as in Fig.~\ref{fig:switch}(b), or by the pure exclusive
binding arrangements of Fig.~\ref{fig:split}.  It appears, however,
that nature has taken advantage of the elegance of the former
scenario: in \Ecoli\ there are a significantly large number of
diverging gene pairs with overlapping operator sites \cite{WtW04_2}.

\section{Discussion}
A genetic switch is inherently stochastic, because of the molecular
character of its components. Our simulations demonstrate that the
stability of a genetic switch can be strongly influenced by its
construction.  If the competing transcription factors mutually exclude
each other at the operator regions, then the switch stability is
markedly enhanced.  Mutual exclusion of competing regulatory molecules
can be obtained by overlapping operons, a network motif that we have
recently identified in a statistical analysis of the gene regulatory
network of \species{E. coli} \cite{WtW04_2}.

The basic conclusion that mutual exclusion of competing regulatory
molecules can strongly enhance the stability of biochemical networks
is robust. Nevertheless, the switch stability is influenced
by phenomena that increase the noise in gene expression.  Such
phenomena include the generation of multiple copies of a transcription
factor from the same mRNA transcript, as well as intrinsic noise
arising from the fluctuations in the numbers of proteins produced from
a transcript.

When expressed as a function of the number of molecules involved in
switching, the switch stability is characterised by a well defined
lifetime $\tau$ which grows exponentially with the mean number,
$\Nmean$, of the transcription factors that are involved in the
switching.  In the regime accessible to direct numerical simulations,
the growth law is well characterised by
$\tau\sim\Nmean{}^{\alpha}\!\exp(b\Nmean)$, where and $\alpha$ and $b$
are parameters.

Whereas we have investigated a number of details that might affect
switch stability, we have left out some considerations which might
additionally be important.  Foremost amongst these is the influence of
cell division and the cell cycle.  Other effects such as fluctuations 
in the availability of RNA polymerase and ribosomes might also have an
influence on switch stability.  Another aspect that should be
investigated is the response of the switch to a perturbation, in other
words how one might toggle a switch by introducing a pulse of some
kind.  We leave all these questions to future work.

The rapid increase of the switch stability with the number of
expressed transcription factors presents a fundamental limitation to
the use of direct simulation to compute the switch lifetime.  This
provokes the question: is there a smarter way to compute the switch
stability for long-lived switches?  Such simulation methods would
generically be useful since the characterisation of rare events
remains a key challenge in computational systems biology \cite{RWA}.
In the field of soft-condensed matter physics, numerical techniques,
such as the reactive flux method~\cite{Bennett77,Chandler78} and
transition path sampling~\cite{BCDG}, have been developed that make it
possible to simulate rare events such as crystal nucleation, protein
folding and chemical reactions. Biochemical networks, however, differ
fundamentally from these problems. In the former problems, the
stationary distribution of states is usually known beforehand. Indeed,
these systems typically obey detailed balance. In contrast, our
genetic circuits, like most biochemical networks, do not satisfy
detailed balance. This means that the stationary distribution of
states is not known {\em a priori}. As a result, numerical techniques
developed to tackle rare events in the field of soft-condensed matter
physics, cannot straightforwardly be transposed to biochemical
networks. Recently, we have developed a new numerical technique,
called Forward Flux Sampling~\cite{Allen04}, that does not rely upon
prior knowledge of the stationary distribution of states.  The scheme
is easily applicable to a wide variety of biochemical switches, such
as that of bacteriophage lambda, and makes it possible to calculate
rates of switching and to identify patwhays for switching. The
technique should therefore lead to a better understanding of these
rare, but important events in biology.
 
\section{Acknowledgements}
We thank Rosalind Allen and Daan Frenkel for useful discussions and
for a critical reading of the manuscript, and Martin Evans for drawing
our attention to the work on driven diffusive systems. We thank Nick
Buchler for drawing our attention to the `split' operator arrangement
shown in Fig. \ref{fig:split}. We are also grateful to the
hospitality of the Kavli Institute for Theoretical Physics in Santa
Barbara, where part of the work was carried out. This research was
supported in part by the National Science Foundation under Grant
No. PHY99-07949. This work is supported by the Amsterdam Centre for
Computational Science (ACCS). The work is part of the research program
of the ``Stichting voor Fundamenteel Onderzoek der Materie (FOM)",
which is financially supported by the ``Nederlandse organisatie voor
Wetenschappelijk Onderzoek (NWO)".

\appendix

\section{Mean field theory for dimerising switches}\label{sec:app1}
Here are the explicit results of the mean field (chemical kinetics)
analysis for the dimerising switches.  Firstly consider the general
switch.  For this case, $\alpha=\beta=\gamma=1$, and
Eqs.~\eqref{fdefeq} and \eqref{gdefeq} simplify to $f=1/(1+y^2)$ and
$g=1/(1+x^2)$.  The discriminant equation $D=0$ solves to
\begin{equation}
y^2=\frac{x^2+1}{3x^2-1}.
\end{equation}  
and requires $x>1/\!\surd{3}\approx0.577$ for solutions.  The cusp of
the wedge in the $(\tmuA,\tmuB)$ plane lies at the point
$\tmuA=\tmuB=1/2$ and corresponds to the point $x=y=1$ on the
line $D=0$.

For the exclusive switch ($\alpha=\beta=1$, $\gamma=0$)
Eqs.~\eqref{fdefeq} and \eqref{gdefeq} give
$f(x,y)=(1+x^2)/(1+x^2+y^2)$ and $g(x,y)=f(y,x)$.  The discriminant
equation $D=0$ now solves to
\begin{equation}
y^2=\frac{\displaystyle{\pm\sqrt{x^8+14x^6+25x^4-24x^2}-x^4-5x^2+2}}
{\displaystyle{2(x^2-1)}}
\end{equation}
This requires $x>\!\surd(3\surd{17}-11)/\!\surd{2}\approx0.827$ for
solutions, at which point $y=\!\surd{3}\approx1.73$. The curve for
$x>1$ corresponds to the positive sign; one can show that the curve
for $x<1$ can be generated by reflecting the curve for $x>1$ through
the line $x=y$ (the curve passes through $x=y=1$).  Also, $y\to1$ as
$x\to\infty$.  The cusp of the wedge in the $(\tmuA,\tmuB)$ plane lies
at the point $\tmuA=\tmuB=2/3$, again corresponding to the point
$x=y=1$ on the line $D=0$.

Finally consider the partially co-operative switch.  This has a much
reduced range of switching behaviour and the corresponding lines have
not been shown in Fig.~\ref{fig:mftheory}.  Consider the case where
$\OO\TFB_2$ is disallowed.  In this case ($\alpha=\gamma=1$,
$\beta=0$) Eqs.~\eqref{fdefeq} and \eqref{gdefeq} give
$f=(1+x^2)/(1+x^2+x^2y^2)$ and $g=1/(1+x^2+x^2y^2)$.  The discriminant
equation $D=0$ solves to
\begin{equation}
y^2=\frac{(x^2+1)^2}{x^2(x^2-5)}
\end{equation}
and requires $x>\!\surd{5}\approx2.24$ to have a solution.  The cusp
of the wedge in the $(\tmuA,\tmuB)$ plane corresponds to the
point $x=\!\surd(13+\!\surd{129})/\!\surd{2}\approx 3.49$ and
$y=\!\surd(51+13\surd{129})/10\approx1.41$, and lies at $\tmuA =
\!\surd(491-43\surd{129})/16 \approx 0.101$ and $\tmuB =
\!\surd(1591\surd{129}-18057)/192 \approx 0.0190$.

\begin{figure*}
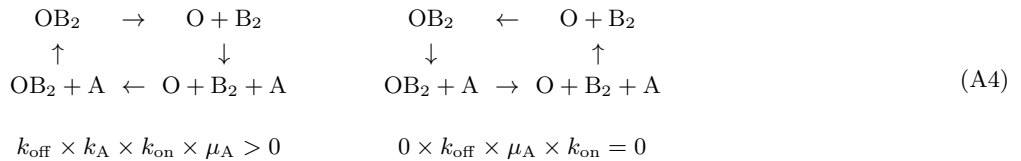

\begin{equation}
\begin{array}{ccc}
\begin{array}{ccc}
\OO\TFB_2 & \to & \OO+\TFB_2 \\
\uparrow & & \downarrow \\
\OO\TFB_2+\TFA & \leftarrow & \OO+\TFB_2+\TFA
\end{array} & \hspace{2em} &
\begin{array}{ccc}
\OO\TFB_2 & \leftarrow & \OO+\TFB_2 \\
\downarrow & & \uparrow \\
\OO\TFB_2+\TFA & \to & \OO+\TFB_2+\TFA
\end{array}\\[24pt]
\koff\times \kA\times\kon\times\muA>0 & &
0\times\koff\times\muA\times\kon=0
\end{array}
\end{equation}
\caption{The fact that the products of the transition rates are
different for the two cycles proves that the corresponding master
equation cannot obey detailed balance.  In both cases, the starting
point is the state $\OO\TFB_2$.\label{fig:cycles}
}
\end{figure*}

We now describe the mean field fixed points for dimerising exclusive
and general switches along the symmetry line $\muA=\muB=\mu$ and
$\kA=\kB=k$.  We assume there is one copy of the genome in the cell.
The reduced degradation rate is
\begin{equation}
\tmu=\frac{\Vcell}{\sqrt{\Kb\Kd}}\,\frac{\mu}{k}.
\end{equation}
For the general switch, Eqs.~\eqref{mfeqs} are
\begin{equation}
\frac{1}{1+y^2} = \tmu x,\quad \frac{1}{1+x^2} = \tmu y.
\end{equation}
The stable fixed points are
\begin{equation}
(x,y)=\frac{\displaystyle{1 \pm \sqrt{1 - 4\tmu^2}}}{2\tmu},
\end{equation}
where $x>y$ if the positive sign is taken for $x$.  We see that
$\tmu<1/2$ is required as found earlier (see Fig.~\ref{fig:mftheory}).
The unstable fixed point is
\begin{equation}
x = y = \frac{u - 12\tmu^2 / u}{6\tmu}
\end{equation}
where
\begin{equation}
u^3 = 108\tmu^2 + 12\tmu^2\sqrt{81 + 12\tmu^2}.
\end{equation}
For the exclusive switch, Eqs.~\eqref{mfeqs} are
\begin{equation}
\frac{1+x^2}{1+x^2+y^2} = \tmu x,\quad
\frac{1+y^2}{1+x^2+y^2} = \tmu y.
\end{equation}
The stable fixed points are
\begin{equation}
(x,y)=\frac{\displaystyle{
[1 - 2\tmu^2 + \sqrt{1 + 4\tmu^2} \pm \Delta]^{1/2}}}
{\displaystyle{2\tmu}}
\end{equation}
where $x>y$ if the positive sign is taken for $x$, and
\begin{equation}
\Delta^2 = 2 - 12\tmu^4 + 2(1 - 2\tmu^2)\sqrt{1 + 4\tmu^2}.
\end{equation}
One can check that $\Delta^2>0$ requires $\tmu<2/3$, as found
earlier.  The unstable fixed point is
\begin{equation}
x = y = \frac{1 + v + (1 - 6\tmu^2) / v}{6\tmu}
\end{equation}
where
\begin{equation}
v^3 = 1 + 45\tmu^2 + 3\tmu\,\sqrt{12 + 213\tmu^2 + 24\tmu^4}
\end{equation}
An expression for the total number of molecules of $\TFA$ in the cell
which applies to both the general ($\gamma=1$) and exclusive
($\gamma=0$) switch is given by
\begin{equation}
\begin{array}{ll}
\NA &= \num{\TFA}+2\,\num{\TFA_2}+2\,\num{\OO\TFA_2}\\[6pt]
&\displaystyle{
= \frac{x\Vcell}{\sqrt{\Kb\Kd}} + \frac{2x^2\Vcell}{\Kb}
+ \frac{2(x^2+\gamma x^2y^2)}{1+x^2+y^2+\gamma x^2y^2}}
\end{array}.\label{naeq}
\end{equation}
The total number of molecules of $\TFB$ is given by the same
expression with $x$ and $y$ interchanged.  This expression, together
with the results above for the locations of the fixed points, is used
to fill in the mean field rows in Table~\ref{tab:features}.
For both kinds of switch, $x=y=1$ at the smallest value of $\tmu$ for
which bistability occurs.  Therefore at this point,
\begin{equation}
\NA=\NB = \frac{\Vcell}{\sqrt{\Kb\Kd}} + \frac{2\Vcell}{\Kb}
+ 2\frac{1+\gamma}{3+\gamma}.\label{nceq}
\end{equation}
We finally consider briefly the limit $\tmu\to0$, which is the limit
of high expression and low degradation rates.  In this limit, the
stable fixed points for both exclusive and general switches converge
with $x\sim 1/\tmu$ and $y\sim\tmu$ (or \latin{vice versa}).  However
the behavior of the unstable fixed point is $x=y$ and
$x\sim\tmu^{-1/3}$ for the general switch, and $x=y$, and $x\sim
1/(2\tmu)$ for the exclusive switch.

\section{Lack of detailed balance}\label{sec:app2}
We prove that the chemical master equation for the set of reactions in
Eqs.~\eqref{chemeqs} cannot satisfy detailed balance.  Recall that
detailed balance implies that 
\begin{equation}
W(1\to2)\,\PS(1)=W(2\to1)\,\PS(2)
\end{equation}
where `1' and `2' indicate any two states, $W(1\to2)$ and $W(2\to1)$
are the forward and backward transition rates between the states, and
$\PS$ is the steady-state probability distribution \cite{kampenbook}.
If detailed balance holds, $\PS$ is unique and all other probability
distributions move towards $\PS$.  For many problems, $\PS$ is known
(eg, it is the Gibbs-Boltzmann distribution).  The problem is then to
find a set of $W$ which satisfies detailed balance, thus guaranteeing
that the dynamics has the correct equilibrium distribution.  This is
the case for Metropolis Monte-Carlo schemes for example.  
For the Gillespie algorithm or the chemical master equation however,
the $W$ are prescribed, and we do not know necessarily know $\PS$.

Without prior knowledge of $\PS$, it would seem difficult to
determine from $W$ alone whether a system obeys detailed balance
or not.  However Mukamel describes a useful test \cite{mukamel}.
Consider any cycle of states $\{1,2,3,\dots,k,1\}$.  A necessary
and sufficient condition for the existence of detailed balance is that
\begin{equation}
\begin{array}{l}
W(1\to2)\,W(2\to3)\dots W(k\to1)\\
\quad=W(1\to k)\,W(k\to k-1)\dots W(2\to1)
\end{array}
\end{equation}
for all such cycles.  To prove that a system does not obey detailed
balance, it suffices to find a single counter-example.

For Eqs.~\eqref{chemeqs}, we can consider the cycles shown in
Fig.~\ref{fig:cycles}. These cycles are mutual inverses, yet the
corresponding products of transition rates are clearly different.  
This proves that the master equation
corresponding to Eqs.~\eqref{chemeqs} does not obey detailed balance.

It is clear that the same argument can be deployed whenever the
appearance of a molecule in the system depends on the state of binding
of some other molecules.  This is very common, for instance it applies
whenever one has gene regulation by a transcription factor.  Thus the
absence of detailed balance would appear to be a generic feature of
any realistic biochemical reaction network.


\begin{thebibliography}{33}
\bibitem[{not()}]{note4}
\bibinfo{note}{Our analysis also improves on Ref.~\onlinecite{WtW04_1} in a
  couple of ways. Firstly, we now define $P(\NA,\NB)$ in Fig.~\ref{fig:densmap}   to be a true time average, proportional to the total time the system spends
  in states with given values of $\NA$ and $\NB$, rather than simply
  proportional to the number of times the system visits states with given
  values of $\NA$ and $\NB$. Secondly, we now calculate the switch lifetime
  from a correlation function method described in the main text, rather than by   analysing the statistics of zero-crossings of the function $\NA-\NB$. The
  zero-crossing method underestimates the lifetime because it detects not only
  true barrier crossing events, but also events where the system briefly visits   states with $\NA\approx\NB$. As the switch lifetime gets longer though, the
  discrepancy becomes less important~\cite{Allen04}.}
\end{thebibliography}
\end{document}